\documentclass[final,5p,times,twocolumn]{elsarticle}
\usepackage{amssymb}
\usepackage{lineno}
\usepackage[utf8]{inputenc}
\usepackage[T1]{fontenc}
\usepackage{epsfig}
\usepackage{amsmath}
\usepackage{amsfonts}
\usepackage{amssymb}
\usepackage{color}
\usepackage{graphicx}
\usepackage{url,hyperref}
\usepackage{latexsym}
\usepackage{longtable}
\usepackage{float}
\usepackage{tikz}
\usepackage{epstopdf}
\usepackage{url}
\usepackage{caption}

\begin{document}
%-------------------------------------------------------------------------------------------------
\begin{frontmatter}
%-------------------------------------------------------------------------------------------------
\title{{\mbox{\small
{}}}\textbf{\ On Hypergraph Representation of Multipartite Quantum Systems}}
\author{Adil Belhaj $^{1}$}
\author{Salah Eddine Ennadifi $^{2}$}
%\author {  \corref {mycorrespondingauthor}}
\cortext[mycorrespondingauthor] {Corresponding author}
\ead{ennadifis@gmail.com}
\author{Zakariae Ennadifi $^{3}$}
\address{
$^{1}$ {ESMaR, Faculty of Science, Mohammed V University in Rabat, Rabat, Morocco.}\\
$^{2}$ {LHEP-MS, Faculty of Science, Mohammed V University in Rabat, Rabat, Morocco.}
\\ $^{3}$ {Faculty of Science, Mohammed V University in Rabat, Rabat, Morocco.}  }

%-------------------------------------------------------------------------------------------------
\begin{abstract}
Borrowing ideas from the  link between   Calabi-Yau singularities and toric geometry in string theory compactifications,   a  hypergraph framework for multipartite quantum systems, 
extending graph representations to higher-order quantum correlations, is
investigated. By implementing  hyperedges in many-body Hamiltonian
interactions, an interplay between hypergraph states and quantum state
dynamics is established. The corresponding  Hamiltonian  hypergraph reveals
the hierarchy of all possible multipartite interactions. However, the  decoherence
effects impose constraints on the realization of large hypergraph states.
The findings  show    the importance of long coherence times and strong many-body
couplings for the realization of large hypergraph states.
\end{abstract}
%-------------------------------------------------------------------------------------------------

%-------------------------------------------------------------------------------------------------
%-------------------------------------------------------------------------------------------------
\end{frontmatter}
%

%-------------------------------------------------------------------------------------------------

\section{Introduction}

Quantum mechanics (QM) is universally recognized as a highly effective fundamental framework  which accurately describes physical phenomena at the atomic scales including  the subatomic one. Precisely, it serves as the cornerstone for all
domains of quantum physics, spanning both rational  sciences and  empirical 
technological applications \cite{1,2,3}. Actually, the manipulation of
multipartite quantum systems lies at the center of modern quantum
information science, with deep implications for quantum communication,
quantum computation, and the comprehension of nonlocal correlations \cite%
{4,5,6}. As the  number $n$  of the quantum system 
increases, however, the characterization of quantum correlations becomes
progressively intricate, especially when  the genuine many-body interactions are
involved. This  has prompted the search for mathematical frameworks capable
to address complex correlation structures while remaining closely
linked to their physical applications.

Hypegraphs, being a generalization of ordinary graphs which 
constitute one of the most successful tools for describing multipartite
quantum states \cite{7,8,9,10}, have been introduced due to the graph theory
limits  describing higher-order interactions involving more than two
qubits. Such a limitation has led to the introduction of hypergraph states
exhibiting a richer correlation structure and more intricate local unitary
operations \cite{8,11,12}. In these hypergraph state secanrios, the  edges are replaced by the 
hyperedges connecting an arbitrary number of vertices, thereby extending
controlled-phase operations to genuine multi-qubit interactions. While the
mathematical structure of hypergraph states is now well established,  their representations have not been   explored in physics issues \cite{13,14,15}.
In particular, a unified framework linking the combinatorial description of
hypergraphs to the underlying Hamiltonian dynamics  which  govern their
generation has not yet been established.  Such a connection is crucial  in order  to 
understand either how abstract hypergraph structures emanate from realistic
many-body interactions, or  how they can be embedded via continuous Schr\"{o}%
dinger evolution. This approach also makes it possible to determine which physical constraints decisively limit their
scalability on quantum platforms \cite{8,16,17}.

\par  Inspired by certain techniques developed in toric geometry within the framework of string theory, we aim to contribute to these activities  by exploring a hypergraph representation of
multipartite quantum models.   With the help of  a
geometric representation of multipartite quantum systems, we  provide a construction
of hypergraph states  $\left\vert \mathcal{H}_{n}\right\rangle $  ($%
n\geqslant 3$) of qubit systems under generalized controlled-$Z$ operations.
We then develop a Hamiltonian formulation of the quantum states of hypergraphs 
in which the $k_\alpha$-hyperedges, associated with generalized operations $C^{k_\alpha}Z_{k_\alpha}$,
directly correspond to a continuous physical evolution over time $U^{\left(
k_\alpha \right) }\left( \Delta t\right) $, establishing a direct bridge 
between hypergraph theory, quantum circuits, and continuous Schrödinger dynamics.  By  examining  the resulting $N$-body Hamiltonian, $\widehat{H}^{\left(
k_\alpha\right) }$, we  investigate the resulting multiparticle interactions, as well as the behavior
of the effective coupling strength $J^\alpha_{eff}=\frac{J^{\left( k_\alpha\right) }}{2^{k_\alpha}%
} $ as a function of the order of the hyperedge $k_\alpha$ and the decoherence effect.

The organization of this paper is as follows. In section 2,  we   expose  a hypergraph representation of  qubit  systems. In section 3, we explore   hypergraph states of such quantum  systems. In section 4,  we present a Hamiltonian  description. 
The last section is devoted to conclusions and open questions.

\section{ Qubit geometric systems using  hypergraph  concepts}
In this section, we build a  hypergraph representation of qubit systems. To show how this can be achieved, we  first consider lower dimensional cases. Then, we  provide the generic description. 
\subsection{One-qubit geometry}
To start,  we  reconsider the 
leading model associated with one qubit. It is recalled that one qubit,
describing a physical system with two states, can be viewed  as a  fundamental 
element in the main discussion of the present task where a novel graph
representation of qubit systems will be established. Roughly,  the  one qubit is
a generic state of a 2-dimensional Hilbert space. Using the Dirac notation,
it can be expressed as follows

\begin{equation}
|\psi >=\sum_{\ell=1}^{2}a_{\ell}|\ell>  \label{1}
\end{equation}%
where $a_{\ell}$ are complex coefficients satisfying the following  normalization
condition 
\begin{equation}
|a_{1}|^{2}+|a_{2}|^{2}=1  \label{2}
\end{equation}%
associated with the probability of one qubit state measurement.
Mathematically, this real algebraic equation (\ref{2}) defines  the unit
sphere $\mathbb{S}^{3}\subset \mathbb{C}^{2}$. Using the phase transformation, however,  this describes a complex geometry
called one dimensional projective space denoted by $\mathbb{CP}^{1}$ defined
usually by the following scale relation 
\begin{equation}
\left( a_{1},a_{2}\right) \sim \lambda \left( a_{1},a_{2}\right)   \label{3}
\end{equation}%
where $\lambda $ is a non-zero complex number. It is well-known that this
complex space is diffeomorphic to a two-dimensional real sphere supported by
quotiening $\mathbb{S}^{3}$  by such a  global phase 
\begin{equation}
\frac{\mathbb{S}^{3}}{U\left( 1\right) }\cong \mathbb{CP}^{1}\cong \mathbb{S}%
^{2}  \label{4}
\end{equation}%
called, in quantum information theory paradigm, the Bloch sphere $\mathbb{S}%
^{2}$ encoding all pure one-qubit states \cite{0100,0101,0102}. This identification is realized by
the following Dirac representation 
\begin{equation}
|\psi >=\cos \left( \frac{\theta }{2}\right) |1>+e^{i\phi }\sin   \left(\frac{%
\theta }{2}\right)|2>  \label{5}
\end{equation}%
where $\theta $ and $\phi $ are the usual spherical variables. It comes that
this geometry may be regarded as a circle $\mathbb{S}^{1}$ fibration over an
interval $\left[ 0,\pi \right] $,  where  the $\mathbb{S}^{1}$ fiber
degenerates  at the two endpoints \cite{100}. Motivated by toric  graph  techniques
used in string theory including D-brane physics, this geometry can be
represented by two vertices and one ordinary  edge \cite{101,102}. It has been suggested that
toric geometry exhibits a nice tool called mirror symmetry playing a crucial
role in the string theory compactification on the Calabi-Yau spaces \cite{103}. In type II superstrings, this application of toric geometry converts the
toric geometry graph of $ \mathbb{S}^{2}$ in type IIA to a vertex in type IIB strings
considered as a powerful procedure in the geometric engineering method of
gauge theory models \cite{103,104}. This method has been extensively exploited in
Calabi-Yau type II superstring compactifications with certain  singularities
classified by Lie algebras which have been approached in various contexts
including M and F-theories\cite{101,103}. This englobes the simply and non simply laced
symmetries known by ADE and BCFG Lie algebras, respectively \cite{105}. In type IIB mirror
geometry,  rouphly, the Bloch sphere $\mathbb{S}^{2}$ can be represented by a vertex 
\begin{equation}
\mbox{qubit}\rightarrow \mbox{Bloch sphere}\rightarrow \mbox{vertex}.
\label{6}
\end{equation}

\subsection{$n$-qubit  geometric systems}

 Having established the leading graph of one qubit, we move to elaborate  the hypergraph
realization that we are after, i.e., a pair $\mathcal{H}_{n}=\left(
V,E\right) $ where $V=\left\{ v_{i}, i=1,\ldots,  n\right\} $ is a finite set
of vertices and $E=\left\{ e_{\alpha}^{\left( k_{\alpha}\right) }, \alpha=1,\ldots, m\right\} $
is the set of hyperedges \cite{7,8}. The numbers $n$, $m$, and $k_\alpha$
are the number of qubits, the number of hyperedges, and the hyperedge order
(cardinality  $\left\vert e_{\alpha}^{\left( k_{\alpha}\right) }\right\vert =k_{\alpha}$
with $k_{\alpha}\leqslant n$), respectively. The corresponding features of such an
algebraic structure will be considered in what follows. Indeed, we consider $%
n$ qubit systems associated with $2^{n}$-dimensional complex Hilbert space $%
\left( \mathbb{C}^{2}\right) ^{\otimes n}$ where a generic state can be
expressed as follows 
\begin{equation}
|\psi >=\sum_{\ell=1}^{2^{n}}a_{\ell}|\ell> \label{7}.
\end{equation}%
Here, $|\ell>$ denotes  a  basis of the corresponding vector space. Now, the complex coefficients $a_{\ell}$ are subject to
the condition 
\begin{equation}
\sum_{\ell=1}^{2^{n}}|a_{\ell}|^{2}=1.  \label{8}
\end{equation}%
A first sight, this  equation describes  a $(2^{n}-1)$-dimensional complex space $%
\mathbb{CP}^{2^{n}-1}$. In order to keep the Bloch sphere representation,
however, we will interpret  this space via real  geometries  in terms of Bloch spheres
associated with individual qubits. In this real geometry language, the corresponding 
dimension is 
\begin{equation}
\dim _{\mathbb{R}}\;\mathcal{S}=2(2^{n}-1) \label{9}
\end{equation}%
where $\mathcal{S}$ is a real space being equivalent to $\mathbb{CP}%
^{2^{n}-1}$ using physical arguments. After a close examination of this
dimension identification, we propose to model $\mathcal{S}$ in terms of a
collection of $(2^{n}-1)$ of two-dimensional real spheres. In this
collection, one has $n$ Bloch spheres corresponding to $n$ qubits according
to the above  scheme  describing the quantum spaces without 
intersections and extra spheres encoding the quantum entanglement space. In
this way, we can split $2^{n}-1$ as follows 
\begin{equation}
2^{n}-1=n+m  \label{10}
\end{equation}%
where one has used $m=2^{n}-n-1$ describing the number of such extra
spheres. This will be supported by standard combinatorics later on. These
extra spheres could play a crucial role in the communication between $n$
qubits. Inspired by graph theory activities in quantum information, they
should be represented by edges $e_{\alpha}^{\left( k_{\alpha}\right) }$, where one has $\alpha=1,\ldots, m $.  Supported by this examination, we can represent $n$ qubits by a hypergraph in two
dimensions with polygon configurations. In this novel graph representation, $%
n$ qubit systems can be represented by $n$ Bloch spheres and $m$
communication spheres producing the quantum entanglement space. Borrowing
ideas from toric geometry in type II superstrings  \cite{103,104}, we replace $n$ qubits by
a hypergraph involving $n$ vertices and $m$ hyperedges $e_{\alpha}^{\left(
k_{\alpha}\right) }$ of orders $k_{\alpha}$,  according to  a  $n$-polygon  being  polygon with $n$ sides.  To see how this works, we shall start
with $n\geqslant 2$ since the case $n=1$ obviously recovers the scheme (\ref{6}) where the hypergraph reduces to a simple one involving only one
vertex $v_{1}$,  where one has $m=0$. Thus, the model $n=2$  describes 
2 qubits corresponding to 4-dimensional Hilbert space. In this situation, it
can be represented by two vertices $v_{1},v_{2},$  and one 2-edge $%
e_{1}^{\left( 2\right) }$ being an ordinary graph representation. This
ordinary graph represents the space $\mathcal{S}$ involving two Bloch
spheres and one communication sphere in the quantum entanglement space side, as illustrated in figure 1.

\begin{figure}[htbp]
    \centering
    \tikzset{every picture/.style={line width=0.65pt}} %set default line width to 0.75pt        

\begin{tikzpicture}[x=0.65pt,y=0.65pt,yscale=-1,xscale=1]
%uncomment if require: \path (0,270); %set diagram left start at 0, and has height of 270

%Straight Lines [id:da14296361178406558]
\draw    (292,122) -- (416.92,122.44) ;
%Shape: Circle [id:dp7497977797745833]
\draw   (242,122) .. controls (242,108.19) and (253.19,97) .. (267,97) .. controls (280.81,97) and (292,108.19) .. (292,122) .. controls (292,135.81) and (280.81,147) .. (267,147) .. controls (253.19,147) and (242,135.81) .. (242,122) -- cycle ;
%Shape: Circle [id:dp8174401489286325]
\draw  [fill={rgb, 255:red, 0; green, 0; blue, 0 }  ,fill opacity=1 ] (351.81,122.22) .. controls (351.81,120.75) and (352.99,119.57) .. (354.46,119.57) .. controls (355.93,119.57) and (357.11,120.75) .. (357.11,122.22) .. controls (357.11,123.69) and (355.93,124.87) .. (354.46,124.87) .. controls (352.99,124.87) and (351.81,123.69) .. (351.81,122.22) -- cycle ;
%Shape: Circle [id:dp734049830943662]
\draw   (416.6,122) .. controls (416.6,108.19) and (427.79,97) .. (441.6,97) .. controls (455.41,97) and (466.6,108.19) .. (466.6,122) .. controls (466.6,135.81) and (455.41,147) .. (441.6,147) .. controls (427.79,147) and (416.6,135.81) .. (416.6,122) -- cycle ;

% Text Node
\draw (258,115) node [anchor=north west][inner sep=0.75pt]  [font=\Large] [align=left] {{\large q\textsubscript{1}}};
% Text Node
\draw (432,115) node [anchor=north west][inner sep=0.75pt]  [font=\Large] [align=left] {{\large q\textsubscript{2}}};

\end{tikzpicture}
    \caption{ Graphic representation of  2-qubits.}
    \label{fig:my-diagram}
\end{figure}
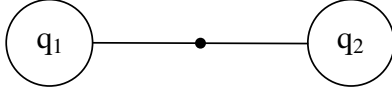

It is easy to show, by counting all possible edge configurations, that the
number of graph states is $2^{C_{n}^{2}}$, where $C_{n}^{2}$ is the binomial
coefficient $n$ choose $2$.

\section{\textbf{Hypergraph state description  of $n$-qubit systems}}
In this section, we investigate the  hypergraph states of $n$-qubit systems using the above concepts.

\subsection{Hypergraph representation}

In the part,  the  hypergraph notation $\mathcal{H}_{n}=\left( V,E\right) $ will be used
for $n\geq 3$ according to \cite{7,8}. Indeed, the case $n=3$ can be represented by $%
\mathcal{H}_{3}$ as a 3-polygon with 3 vertices $v_{1},v_{2},v_{3}$ and four
hyperedges $e_{1}^{\left( k_{1}\right) },e_{2}^{\left( k_{2}\right)
},e_{3}^{\left( k_{3}\right) },e_{4}^{\left( k_{4}\right) }$. This
hypergraph involves three vertices corresponding to 3-orthogonal Bloch
spheres defining the associated quantum space without the intersections and
four hyperedges describing the communication spheres which could be
exploited in the quantum entanglement geometry. This  interplay  can be illustrated in figure
2.
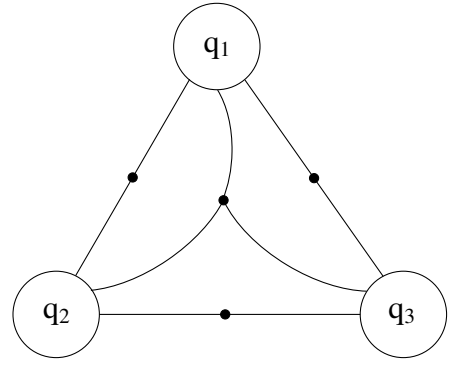
\begin{figure}[htbp]
    \centering
   \begin{tikzpicture}[x=0.65pt,y=0.65pt,yscale=-1,xscale=1]
%uncomment if require: \path (0,270); %set diagram left start at 0, and has height of 270

%Straight Lines [id:da14296361178406558]
\draw    (251.28,193.56) -- (317.28,79.96) ;
%Shape: Circle [id:dp7497977797745833]
\draw   (215,216) .. controls (215,202.19) and (226.19,191) .. (240,191) .. controls (253.81,191) and (265,202.19) .. (265,216) .. controls (265,229.81) and (253.81,241) .. (240,241) .. controls (226.19,241) and (215,229.81) .. (215,216) -- cycle ;
%Shape: Circle [id:dp7339636053906097]
\draw   (416,216) .. controls (416,202.19) and (427.19,191) .. (441,191) .. controls (454.81,191) and (466,202.19) .. (466,216) .. controls (466,229.81) and (454.81,241) .. (441,241) .. controls (427.19,241) and (416,229.81) .. (416,216) -- cycle ;
%Straight Lines [id:da26035906411933896]
\draw    (265,216) -- (416,216) ;
%Shape: Circle [id:dp8174401489286325]
\draw  [fill={rgb, 255:red, 0; green, 0; blue, 0 }  ,fill opacity=1 ] (281.63,136.76) .. controls (281.63,135.29) and (282.81,134.11) .. (284.28,134.11) .. controls (285.75,134.11) and (286.93,135.29) .. (286.93,136.76) .. controls (286.93,138.23) and (285.75,139.41) .. (284.28,139.41) .. controls (282.81,139.41) and (281.63,138.23) .. (281.63,136.76) -- cycle ;
%Shape: Circle [id:dp734049830943662]
\draw   (308,61) .. controls (308,47.19) and (319.19,36) .. (333,36) .. controls (346.81,36) and (358,47.19) .. (358,61) .. controls (358,74.81) and (346.81,86) .. (333,86) .. controls (319.19,86) and (308,74.81) .. (308,61) -- cycle ;
%Straight Lines [id:da013484239994840164]
\draw    (429.28,193.96) -- (349.28,80.36) ;
%Shape: Circle [id:dp18732615706392852]
\draw  [fill={rgb, 255:red, 0; green, 0; blue, 0 }  ,fill opacity=1 ] (335.19,216) .. controls (335.19,214.53) and (336.38,213.35) .. (337.85,213.35) .. controls (339.31,213.35) and (340.5,214.53) .. (340.5,216) .. controls (340.5,217.47) and (339.31,218.65) .. (337.85,218.65) .. controls (336.38,218.65) and (335.19,217.47) .. (335.19,216) -- cycle ;
%Shape: Circle [id:dp2163138772893325]
\draw  [fill={rgb, 255:red, 0; green, 0; blue, 0 }  ,fill opacity=1 ] (386.63,137.16) .. controls (386.63,135.69) and (387.81,134.51) .. (389.28,134.51) .. controls (390.75,134.51) and (391.93,135.69) .. (391.93,137.16) .. controls (391.93,138.63) and (390.75,139.81) .. (389.28,139.81) .. controls (387.81,139.81) and (386.63,138.63) .. (386.63,137.16) -- cycle ;
%Shape: Circle [id:dp1350967083897665]
\draw  [fill={rgb, 255:red, 0; green, 0; blue, 0 }  ,fill opacity=1 ] (334.23,149.96) .. controls (334.23,148.49) and (335.41,147.31) .. (336.88,147.31) .. controls (338.35,147.31) and (339.53,148.49) .. (339.53,149.96) .. controls (339.53,151.43) and (338.35,152.61) .. (336.88,152.61) .. controls (335.41,152.61) and (334.23,151.43) .. (334.23,149.96) -- cycle ;
%Shape: Arc [id:dp2709018091919625]
\draw  [draw opacity=0] (260.1,202.07) .. controls (273.53,200.74) and (289.41,194.77) .. (304.47,184.5) .. controls (319.39,174.33) and (330.69,161.88) .. (336.88,149.96) -- (284.71,155.51) -- cycle ; \draw   (260.1,202.07) .. controls (273.53,200.74) and (289.41,194.77) .. (304.47,184.5) .. controls (319.39,174.33) and (330.69,161.88) .. (336.88,149.96) ;  
%Shape: Arc [id:dp43032987805732703]
\draw  [draw opacity=0] (337.12,150.54) .. controls (343.37,163.39) and (356.31,176.98) .. (373.65,187.44) .. controls (389.58,197.05) and (406.1,202.15) .. (419.75,202.57) -- (391.78,157.4) -- cycle ; \draw   (337.12,150.54) .. controls (343.37,163.39) and (356.31,176.98) .. (373.65,187.44) .. controls (389.58,197.05) and (406.1,202.15) .. (419.75,202.57) ;  
%Shape: Arc [id:dp21026231191559752]
\draw  [draw opacity=0] (336.45,149.35) .. controls (339.64,141.69) and (341.61,132.22) .. (341.79,121.96) .. controls (342.04,107.7) and (338.79,94.88) .. (333.46,86.2) -- (316.28,121.51) -- cycle ; \draw   (336.45,149.35) .. controls (339.64,141.69) and (341.61,132.22) .. (341.79,121.96) .. controls (342.04,107.7) and (338.79,94.88) .. (333.46,86.2) ;  

% Text Node
\draw (231,208) node [anchor=north west][inner sep=0.75pt]  [font=\Large] [align=left] {{\large q\textsubscript{2}}};
% Text Node
\draw (431,208) node [anchor=north west][inner sep=0.75pt]  [font=\Large] [align=left] {{\large q\textsubscript{3}}};
% Text Node
\draw (324,54) node [anchor=north west][inner sep=0.75pt]  [font=\Large] [align=left] {{\large q\textsubscript{1}}};

\end{tikzpicture}
    \caption{Hypergraph representation of 3-qubits.}
    \label{fig:my-diagram}
\end{figure}

In this graph representation,  the  hyperedges are split into two categories. The first one contains 3
ordinary edges  being 2-edges $e_{1}^{\left( 2\right) },e_{2}^{\left(
2\right) }$ and $e_{3}^{\left( 2\right) }$, where one  has  $k_1= k_2= k_3=2$.  The second  one, however,   involves only one
3-edge $e_{4}^{\left( 3\right) }$, where  one has   $k_3=4$.

This analysis can be extended to $n$%
qubits. In this way, they can be represented by a hypergraph as a $n$ 
-polygon with $n$ vertices $v_{i}$, ($i=1,\ldots, n$)  and $m$ hyperedges $%
e_{\alpha}^{\left( k_{\alpha}\right) }$ of oders $k_{\alpha}$. This graphic representation  involves $n-1$
categories of $k_\alpha$-edges. Concretely, the number of $k_\alpha$-edges $e_{\alpha}^{\left(
k_\alpha \right) }$ e.g., $|E_{k_\alpha}|$   in each category is
\begin{equation} 
|E_{k\alpha}|=C_{n}^{k_\alpha}. \label{11}
\end{equation}
 Thus, the total number of hyperedges  $|E|$  in (\ref{10}) is obtained such as
\begin{equation}
|E| =\sum\limits_{k_\alpha=2}^{n}C_{n}^{k_\alpha}=2^{n}-n-1=m, \quad 
n\geqslant 2.  \label{12}
\end{equation}
In this scenario, the  total number of hypergraph states for $n$ vertices turns out to be $%
2^{\sum\limits_{k_\alpha-=2}^{n}C_{n}^{k_\alpha}}=2^{2^{n}-n-1}.$

\subsection{Hypergraph states}

Provided a mathematical hypergraph $\mathcal{H}_{n}=\left( V,E\right) $, the
corresponding quantum state $\left\vert \mathcal{H}_{n}\right\rangle $ can
be obtained by assigning to each vertex $v_{i}$ belonging  to $ V$ an initialised qubit $%
\left\vert +\right\rangle =\frac{1}{\sqrt{2}}\left( \left\vert
0\right\rangle +\left\vert 1\right\rangle \right) $, and to each hyperedge $%
e_{\alpha}^{\left( k_{\alpha}\right) }$ a controlled-$Z$ operation to be performed
between all connected qubits \cite{10,11,12,13,14,15}. Concretely, \ for a $n
$ qubit system in which $k_\alpha $ qubits $v_{1},v_{2},\ldots, v_{k_\alpha}$ are connected by a 
$k_\alpha$-hyperedge, we have the  following scheme
\begin{eqnarray}
v_{i} &\equiv &\left\vert +\right\rangle _{i}\rightarrow \text{\ }V\equiv
\left\vert +\right\rangle ^{\otimes n}  \label{13} \\
e_{\alpha}^{\left( k_\alpha \right) } &\rightarrow &\widehat{e_\alpha{}^{\left(
k_\alpha \right) }}\equiv C^{k_\alpha}Z_{ v_{1},v_{2},\ldots, v_{k_\alpha}}  \label{14}
\end{eqnarray}
where $\left\vert +\right\rangle ^{\otimes n}$ is the total initial state
and $C^{k_\alpha}Z_{v_{1},v_{1},\ldots,v_{k_\alpha}}$ is the controlled-$Z$ operator\footnote{
By definition  $C^{0}Z=-1, C^{1}Z_{v_{1}}=1$} corresponding to the $k_\alpha$-hyperedge $e_{\alpha}^{\left(
k_{\alpha}\right) }$ connecting $k_\alpha$ qubits associated with  the $k_\alpha$ vertices $%
v_{1},v_{2},\ldots, v_{k_\alpha}$. The gate $C^{k_\alpha}Z_{v_{1},v_{2},\ldots, v_{k_\alpha}}=diag(\overset{%
2^{k_\alpha}-1}{\overbrace{1,1,\ldots,1}},-1)$ acts on the input state $\left\vert
11...1\right\rangle _{v_{1},v_{2},\ldots, v_{k_\alpha}}\equiv \left\vert 1\right\rangle
^{\otimes k_\alpha}$ by introducing a minus sign on the target $k_\alpha$ qubit such as $%
C^{k_\alpha}Z_{v_{1},v_{2},\ldots, v_{k_\alpha}}\left\vert 1\right\rangle ^{\otimes k_\alpha}=\left(
- 1\right) \overset{k_\alpha}{^{\overbrace{1\times \ldots \times 1}}}\left\vert
1\right\rangle ^{\otimes k}=-\left\vert 1\right\rangle ^{\otimes k_\alpha}$. \ More
explicitly, such a unitary gate can be written as an operator using the
identity matrix $I=I_{2^{k_\alpha}\times 2^{k_\alpha}}$ and the projector onto the $%
\left\vert 1\right\rangle ^{\otimes k_\alpha}$ state

\begin{equation}
C^{k_\alpha}Z_{v_{1},v_{2},\ldots, v_{k_\alpha}}=I-2\left\vert 1\right\rangle ^{\otimes
k_\alpha}\left\langle 1\right\vert ^{\otimes k_\alpha}=I-2P_{v_{1},v_{2}, \ldots, v_{k_\alpha}}^{\left(
k_\alpha\right) }=e^{i\pi P_{v_{1},v_{2},\ldots, v_{k_\alpha}}^{\left( k_\alpha \right) }}  \label{15}
\end{equation}%
where the projector $P^{\left( k_\alpha \right) }\equiv \left\vert 1\right\rangle
^{\otimes k_\alpha }\left\langle 1\right\vert ^{\otimes k_\alpha }$, being  a $2^{k_\alpha}\times 2^{k_\alpha}$
matrix, has been used. It represents the multi-body projector onto the shared excited state
of the $k_\alpha$ connected qubits.  Eventually,  we get the quantum state

\begin{eqnarray}
\left\vert \mathcal{H}_{n}\right\rangle&=& \prod\limits_{k_\alpha=1}^{n}\prod\limits_{\left( v_{1},v_{1},\ldots, v_{k_\alpha}\right) 
}e^{i\pi P_{v_{1},v_{2},\ldots,v_{k_\alpha}}^{\left( k_\alpha \right) }}\left\vert
+\right\rangle ^{\otimes n} \nonumber
\\&=&e^{\left( i\pi \left[ \sum\limits_{k_\alpha=1}^{n}%
\sum\limits_{\left( v_{1},v_{1},\ldots, v_{k_\alpha}\right) 
}P_{v_{1},v_{2},.\ldots, v_{k_\alpha}}^{\left( k_\alpha \right) }\right] \right) }\left\vert
+\right\rangle ^{\otimes n}\\\nonumber
\label{16}
\end{eqnarray}
where the product of $\prod\limits_{k_\alpha=1}^{n}$ accounts for different types of
hyperedges $k_\alpha \leqslant n$ in the hypergraph  where  $k_\alpha$-hyperedge connects  $%
k_\alpha$ vertices.  As illustrated in figure 3,  the  following circuit representation shows such a
correspondence scheme for the hpergraph state $\left\vert \mathcal{H}%
_{3}\right\rangle $.

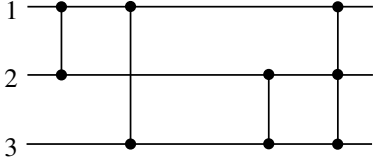
\begin{figure}[htbp]
 \centering  
\tikzset{every picture/.style={line width=0.65pt}} %set default line width to 0.75pt        

\begin{tikzpicture}[x=0.65pt,y=0.65pt,yscale=-1,xscale=1]

%Straight Lines [id:da531038084636223]
\draw    (190.4,90.15) -- (390.4,90.15) ;
%Shape: Circle [id:dp14068713400838195]
\draw  [fill={rgb, 255:red, 0; green, 0; blue, 0 }  ,fill opacity=1 ] (207.38,90.23) .. controls (207.38,88.72) and (208.6,87.51) .. (210.1,87.51) .. controls (211.6,87.51) and (212.82,88.72) .. (212.82,90.23) .. controls (212.82,91.73) and (211.6,92.95) .. (210.1,92.95) .. controls (208.6,92.95) and (207.38,91.73) .. (207.38,90.23) -- cycle ;
%Shape: Circle [id:dp3396437329711791]
\draw  [fill={rgb, 255:red, 0; green, 0; blue, 0 }  ,fill opacity=1 ] (247.38,90.04) .. controls (247.38,88.54) and (248.6,87.32) .. (250.1,87.32) .. controls (251.6,87.32) and (252.82,88.54) .. (252.82,90.04) .. controls (252.82,91.55) and (251.6,92.76) .. (250.1,92.76) .. controls (248.6,92.76) and (247.38,91.55) .. (247.38,90.04) -- cycle ;
%Shape: Circle [id:dp7280025243281718]
\draw  [fill={rgb, 255:red, 0; green, 0; blue, 0 }  ,fill opacity=1 ] (207.38,129.65) .. controls (207.38,128.15) and (208.6,126.93) .. (210.1,126.93) .. controls (211.6,126.93) and (212.82,128.15) .. (212.82,129.65) .. controls (212.82,131.16) and (211.6,132.37) .. (210.1,132.37) .. controls (208.6,132.37) and (207.38,131.16) .. (207.38,129.65) -- cycle ;
%Shape: Circle [id:dp7194532385324162]
\draw  [fill={rgb, 255:red, 0; green, 0; blue, 0 }  ,fill opacity=1 ] (247.38,169.65) .. controls (247.38,168.15) and (248.6,166.93) .. (250.1,166.93) .. controls (251.6,166.93) and (252.82,168.15) .. (252.82,169.65) .. controls (252.82,171.16) and (251.6,172.37) .. (250.1,172.37) .. controls (248.6,172.37) and (247.38,171.16) .. (247.38,169.65) -- cycle ;
%Shape: Circle [id:dp2061239022891539]
\draw  [fill={rgb, 255:red, 0; green, 0; blue, 0 }  ,fill opacity=1 ] (327.38,129.37) .. controls (327.38,127.86) and (328.6,126.65) .. (330.1,126.65) .. controls (331.6,126.65) and (332.82,127.86) .. (332.82,129.37) .. controls (332.82,130.87) and (331.6,132.09) .. (330.1,132.09) .. controls (328.6,132.09) and (327.38,130.87) .. (327.38,129.37) -- cycle ;
%Shape: Circle [id:dp4670755929089312]
\draw  [fill={rgb, 255:red, 0; green, 0; blue, 0 }  ,fill opacity=1 ] (327.38,169.51) .. controls (327.38,168.01) and (328.6,166.79) .. (330.1,166.79) .. controls (331.6,166.79) and (332.82,168.01) .. (332.82,169.51) .. controls (332.82,171.01) and (331.6,172.23) .. (330.1,172.23) .. controls (328.6,172.23) and (327.38,171.01) .. (327.38,169.51) -- cycle ;
%Shape: Circle [id:dp6665628471756834]
\draw  [fill={rgb, 255:red, 0; green, 0; blue, 0 }  ,fill opacity=1 ] (367.38,90.22) .. controls (367.38,88.72) and (368.6,87.5) .. (370.1,87.5) .. controls (371.6,87.5) and (372.82,88.72) .. (372.82,90.22) .. controls (372.82,91.73) and (371.6,92.94) .. (370.1,92.94) .. controls (368.6,92.94) and (367.38,91.73) .. (367.38,90.22) -- cycle ;
%Shape: Circle [id:dp4904226320839259]
\draw  [fill={rgb, 255:red, 0; green, 0; blue, 0 }  ,fill opacity=1 ] (367.28,129.37) .. controls (367.28,127.86) and (368.5,126.65) .. (370,126.65) .. controls (371.5,126.65) and (372.72,127.86) .. (372.72,129.37) .. controls (372.72,130.87) and (371.5,132.09) .. (370,132.09) .. controls (368.5,132.09) and (367.28,130.87) .. (367.28,129.37) -- cycle ;
%Shape: Circle [id:dp5982110805777615]
\draw  [fill={rgb, 255:red, 0; green, 0; blue, 0 }  ,fill opacity=1 ] (367.28,169.73) .. controls (367.28,168.23) and (368.5,167.01) .. (370,167.01) .. controls (371.5,167.01) and (372.72,168.23) .. (372.72,169.73) .. controls (372.72,171.23) and (371.5,172.45) .. (370,172.45) .. controls (368.5,172.45) and (367.28,171.23) .. (367.28,169.73) -- cycle ;
%Straight Lines [id:da3189383067751923]
\draw    (190.4,129.65) -- (390.4,129.65) ;
%Straight Lines [id:da9243065642791637]
\draw    (189.9,169.65) -- (389.9,169.65) ;
%Straight Lines [id:da08186431559678853]
%\draw    (250.4,49.76) -- (250.4,60.78) ;
%\draw   (266.6,49.76) -- (269.8,55.47) -- (266.6,61.18) ;
%Straight Lines [id:da4616789684415147]
\draw    (210.1,90.23) -- (210.1,129.65) ;
%Straight Lines [id:da9115257795460932]
\draw    (250.1,90.04) -- (250.1,169.65) ;
%Straight Lines [id:da8788357176459008]
\draw    (370.1,90.22) -- (370.1,169.83) ;
%Straight Lines [id:da44281553170261945]
\draw    (330.1,129.7) -- (330.1,169.51) ;

% Text Node
\draw (175.9,85.5) node [anchor=north west][inner sep=0.75pt]   [align=left] {{\normalsize 1}};
% Text Node
\draw (175.4,124.5) node [anchor=north west][inner sep=0.75pt]   [align=left] {{\normalsize 2}};
% Text Node
\draw (175.4,164.5) node [anchor=north west][inner sep=0.75pt]   [align=left] {{\normalsize 3}};
% Text Node
%\draw (250.68,49.5) node [anchor=north west][inner sep=0.75pt]   [align=left] {{\footnotesize H}{\tiny 3}};

\end{tikzpicture}
\caption{Hpergraph state $\left\vert \mathcal{H}%
_{3}\right\rangle. $ }
    \label{fig:my-diagram}
\end{figure}

This is a  leading example of the quantum hypergraph state $\left\vert \mathcal{H}%
_{3}\right\rangle $ obtained from the hypergraph $\mathcal{H}_{3}$ for $n=3$. In this way,  the  gates are represented by a line, with  dots indicating
qubits on which they are acting.

\section{Hamiltonian formulation for hypergraph states }

\subsection{Dynamical evolution}

From a dynamical perspective, the $C^{k_\alpha}Z_{v_{1},v_{1},\ldots, v_{k_\alpha}}$ discrete
gates generating the hypergraph state map directly to continuous physical
time evolution \cite{16,17}. In the Schr\"{o}dinger picture, any unitary
gate $U$ is generated by an interaction Hamiltonian $\widehat{H}$ over a
duration $\Delta t$ through $U=e^{-iH\Delta t}$.  Since  the
gates   (\ref{15}) consist entirely of commuting Pauli-$Z$ operators, the entire
hypergraph state  can be mapped directly onto a multi-body
interaction Hamiltonian.  Such an explicit connection bridges the theoretical
graph description of the $n$-qubits state with the realistic physical fields
and couplings. Since the initial state at $t=0$ is
\begin{equation}
|\psi \left( t=0\right) >\equiv \left\vert +\right\rangle ^{\otimes n},
\label{17}
\end{equation}
the continuous evolution over a duration $\Delta t$ yields  \footnote{Natural units  have been considered $\hbar =c=1$.}
\begin{eqnarray}
U\left( \Delta t\right) \left\vert +\right\rangle ^{\otimes n}&=&e^{-i\widehat{%
H}\Delta t}\left\vert +\right\rangle ^{\otimes n}  \nonumber \\&=&e^{-i\left(
\sum\limits_{k_\alpha=1}^{n}\sum\limits_{\left( v_{1},v_{1},\ldots,v_{k_\alpha}\right)}%
\widehat{H}_{v_{1},v_{2},\ldots, v_{k_\alpha}}^{\left( k_\alpha \right) }\right) \Delta
t}\left\vert +\right\rangle ^{\otimes n} \nonumber \\&=&|\psi \left( \Delta t\right)
>\\&=&\left\vert \mathcal{H}_{n}\right\rangle  \label{18} \nonumber
\end{eqnarray}
where $\widehat{H}$ is the total Hamiltonian.  $\widehat{H}%
_{v_{1},v_{2},\ldots, v_{k_\alpha}}^{\left( k_\alpha \right) }$  represents the $k_\alpha$-qubit interaction
Hamiltonian for a single $k_\alpha$-hyperedge. Concretely,  it is written as 
\begin{equation}
\widehat{H}_{v_{1},v_{2},\ldots, v_{k_\alpha}}^{\left( k_\alpha \right) }=-\frac{\pi }{\Delta t ^{\left( k_\alpha \right)}}%
P_{v_{1},v_{2},\ldots, v_{k_\alpha}}^{\left( k_\alpha \right) }  \label{19}
\end{equation}%
Alternatively, this interaction can be re-expresssed in terms of an explicit
physical coupling strength $J^{\left( k_\alpha \right) }$ as
\begin{equation}
\widehat{H}_{v_{1},v_{2},\ldots, v_{k_\alpha}}^{\left( k_\alpha \right) }=-J^{\left( k_\alpha \right)
}P_{v_{1},v_{2},\ldots, v_{k_\alpha}}^{\left( k_\alpha \right) }  \label{20}
\end{equation}%
where  $P_{v_{1},v_{2},\ldots, v_{k_\alpha}}^{\left( k_\alpha \right) } $  is now the multi-body
projector onto the shared excited state of the $k_\alpha$-connected qubits. To
recover the exact target logic of the discrete hypergraph   (\ref{15}), we enforce
a pulsed-evolution protocol under the phase-matching constraint.    Concretely,  we consider 
an interaction duration such that
\begin{equation}
\Delta t^{\left( k_\alpha \right) } J^{\left( k_\alpha \right) }=\pi .  \label{21}
\end{equation}
Under such a gate-generation condition (\ref{21}), the unitary
time-evolution (\ref{18}) operator for a single hyperedge $U^{\left(
k_\alpha \right) }$ evaluates directly to the desired phase gate $%
C^{k_\alpha}Z_{v_{1},v_{2},\ldots, v_{k_\alpha}}$. Indeed, we have
\begin{equation}
U^{\left(  k_\alpha \right) }\left( \frac{\pi }{J^{\left(  k_\alpha \right) }}\right)
\left\vert +\right\rangle ^{\otimes  k_\alpha }=e^{-i\widehat{H}%
_{v_{1},v_{2},\ldots, v_{k_\alpha }}^{\left(  k_\alpha \right) } \Delta t^{\left( k_\alpha \right) }}\left\vert +\right\rangle
^{\otimes k_\alpha }=e^{i\pi P_{v_{1},v_{2},\ldots, v_{k_\alpha }}^{\left( k_\alpha \right) }}\left\vert
+\right\rangle ^{\otimes k_\alpha }.  \label{22}
\end{equation}

\subsection{Multipartite interaction}
To reveal the explicit spin-coupling structure of the system, we expand the
projector $P_{v_{1},v_{2},\ldots, v_{k_\alpha}}^{\left( k_\alpha \right) }$ in the Pauli-$Z$
basis to obtain the corresponding Hamiltonian form
\begin{equation}
\widehat{H}_{v_{1},v_{2},\ldots, v_{k_\alpha}}^{\left( k_\alpha \right) }=-\frac{J^{\left(
k_\alpha \right) }}{2^{k_\alpha}}\left[ I+ \prod \limits_{i=1} ^{k_\alpha}  \left( -1\right) ^{k_\alpha}  Z_{v_ i}\right]
  \label{23}
\end{equation}
where  the Pauli operator
product $Z_{v_{1}}Z_{v_{2}}\ldots Z_{v_{k_{\alpha}}}$ defines the physical
joint-spin operator. Each vertex index $v_{i}$ identifies the precise
target qubit on which the  specific Pauli-$Z$ matrix acts locally. This $k_\alpha$ hyperedge (\ref{23}) Hamiltonian contains all possible interaction orders.
Indeed, ignoring the first term proportional to $I$ representing a constant
residual energy shift affecting the entire system uniformly, the term proportional to $Z_{v_{i}}$    ($k_{\alpha}=1$)  represent an independent energy configurations
acting locally on each individual qubit within the hyperedge, the term ($k_{\alpha}=2
$) proportional to $Z_{v_{i}}Z_{v_{j}}$ correspond to conventional Ising
interactions capable of generating pairwise entanglement. However, the term  proportional to $Z_{v_{1}}Z_{v_{2}}.\ldots Z_{v_{k_\alpha}}$ represents the higher
order interaction that couples all $k_\alpha$ qubits simultaneously responsible for
the creation of genuine multipartite correlations where the hypergraph  description  is relevant. 

\subsection{Decoherence and scalability}

The general expression (\ref{23}) shows that for a constant coupling strength $%
J^{\left( k_\alpha \right) }$ as the hyperedge order $k_\alpha$ grows larger, the
denominator $2^{k}$ progressively weakens the strength of individual
couplings, posing a significant challenge for large-scale physical
engineering. Physically, a weak effective coupling $J^{(k_\alpha)}_{eff}=\frac{J^{\left(
k_\alpha \right) }}{2^{k_\alpha }}$ imposes tough  operational constraints. As the
interaction strength scales down by $2^{k_\alpha}$, the required gate time
\begin{equation}
\Delta t  ^{(k_\alpha)}=\frac{2^{k_\alpha}\pi }{J^{\left( k_\alpha \right) }}  \label{24}
\end{equation}%
increases exponentially. This makes  the protocol highly susceptible to a  standard
environmental decoherence before the hypergraph entanglement can be
finalized. Concretely, during the extended interaction time (\ref{24}) 
uncontrolled coupling to the environment   induces dephasing and decoherence,
progressively suppressing the off-diagonal coherences of the evolving
quantum state and  destroys the fragile evolving quantum superposition as follows 
\begin{equation}
\left\vert \psi \left( t\right) \right\rangle =e^{-i\widehat{H}%
_{v_{1},v_{2},\ldots, v_{k_\alpha }}^{\left( k_\alpha \right) }t}\left\vert +\right\rangle
^{\otimes k_\alpha }, \text{ \ \ \ \ }0<t< \Delta t  ^{(k_\alpha)}. \label{25}
\end{equation}%
This scenario washes  away the emerging multipartite entanglement before the Hamiltonian
evolution can successfully complete the hypergraph state generation.
Consequently, for a fixed coherence time $T_{c}$, there exists a practical
upper bound on the realizable hyperedge order $k_\alpha \leqslant k_\alpha^{\max }$.
Indeed, by imposing $\Delta t  ^{(k_\alpha)}<T_{c}$,  we get
\begin{equation}
k_\alpha ^{\max }=\log _{2}\left( \frac{J^{\left( k_\alpha \right) }T_{c}}{\pi }\right).
\label{26}
\end{equation}%
This emphasizes the need for long coherence times and strong many-body
couplings to realize large hypergraph states.

\section{Concluding remarks}

In this work, we have considered a class of quantum hypergraph states that
naturally generalizes graph states. We have shown that these hypergraph
states, which are associated with mathematical hypergraphs and are generated
by generalized controlled-phase operations acting on multi-qubit hyperedges,
admit a direct Hamiltonian realization in terms of many-body interactions. This has 
provided  a bridge between hypergraph theory and continuous Schr\"{o}dinger
dynamics.  Precisely, we have  revealed  that such a hypergraph state
construction exhibits  a natural Hamiltonian realization in terms of many-body
projector interactions. Within such a realization, the hypergraph structure
translates directly into a hierarchy of multi-qubit couplings, offering a
unified framework connecting combinatorial hypergraphs and continuous Schr%
\"{o}dinger dynamics.  The corresponding Hamiltonian formulation  has  shown 
an intrinsic scalability limitation arising from the exponential reduction
of effective interaction strengths with respect to the hyperedge orders. It has been   remarked that this 
imposes  a fundamental scalability constraint due to decoherence effects.  As a result,  we have demonstrated that such  a scenario 
emphasizes the importance of long coherence times and strong many-body
couplings for realizing large multipartite hypergraph states.

This work comes up with certain open questions.   Since hypergraph states can also serve as a fertile ground for foundational
investigations, it will be of great interest for future studies to ask
whether some results of certain quantum phenomena\ can be dealt with within
such hypergraph  approaches.  Inspired by   toric geometry and string theory descriptions of
qudit systems, it could be possible to implement hypergraphs in  the study of Calabi-Yau  manifolds explored in string theory and related topics to  approach non-trivial singularities   going beyond  the ones classified by  Lie algebras including non simply laced ones.   We hope to report elsewhere on such  open questions in future investigations.

\end{document}